\documentclass[a4paper,fleqn,usenatbib]{mnras}

\usepackage{amsmath}
\usepackage{amsfonts}
\usepackage{amsbsy}
\usepackage{amssymb}
\usepackage{amsbsy}

\usepackage[T1]{fontenc}
\usepackage{ae,aecompl}

\usepackage{graphicx}
\usepackage{gensymb}
\usepackage{float}
\usepackage[utf8]{inputenc}
\usepackage{tensor}
\usepackage{bm}
\usepackage{enumerate}

\title[Spiral waves and vertical circulation]{Spiral density waves and vertical circulation in protoplanetary discs}

\author[Riols and Latter]{
A. Riols,$^{1,2}$ H. Latter $^{1}$
\\
$^{1}$Department of Applied Mathematics and Theoretical Physics, University of Cambridge, Centre for Mathematical Sciences, \\
Wilberforce Road, Cambridge CB3 0WA, UK.\\
$^{2}$ Institut de Plan\'etologie et d'Astrophysique de Grenoble, BP 53 
38041 Grenoble, Cedex 9, FRANCE}

\date{Accepted XXX. Received YYY; in original form ZZZ}

\pubyear{2017}

\begin{document}
\label{firstpage}
\pagerange{\pageref{firstpage}--\pageref{lastpage}}
\maketitle

\begin{abstract}
Spiral density waves dominate several facets of
accretion disc dynamics --- planet-disc interactions and
gravitational instability (GI) most prominently. 
Though they have been examined thoroughly in two-dimensional
simulations, their vertical structures in the non-linear regime are
somewhat unexplored.
 This neglect is unwarranted given that any strong vertical motions
 associated with these waves could profoundly impact dust dynamics,
dust sedimentation, planet formation, and the emissivity
of the disc surface. In this paper we combine linear calculations and
shearing box simulations in order to investigate the vertical structure
of spiral waves for various polytropic stratifications and wave
amplitudes. 
For sub-adiabatic profiles we find that
spiral waves develop a pair of counter-rotating poloidal
rolls.  Particularly strong in the nonlinear regime, these vortical
structures issue from the baroclinicity supported by the 
background vertical entropy gradient. They are also intimately connected to
the disk's g-modes which appear to interact nonlinearly with the density waves. 
Furthermore, we demonstrate that the poloidal rolls
are ubiquitous in gravitoturbulence, emerging in the vicinity of GI
spiral wakes, and potentially
transporting grains off the disk midplane. Other than hindering 
sedimentation and planet formation, this phenomena may bear on
observations of the disk's
scattered infrared luminosity. 
 The vortical features could also impact on the
 turbulent dynamo operating 
 in young protoplanetary discs subject to GI, or possibly even 
 galactic discs.  
\end{abstract}

\begin{keywords}
accretion discs --- turbulence --- protoplanetary discs  --- waves
\end{keywords}



\section{Introduction}
 
Spiral density waves participate in several processes 
controlling the evolution and structure of accretion discs.
They transport appreciable angular momentum, 
while also actively changing
the observable structure of the system. Waves can also serve
as diagnostics, telling us information about the disk and the
perturbers that generate them. 
Spiral waves
have been directly observed in various gaseous protoplanetary 
(PP) disks  \citep[e.g.\ MWC 758,
HD 142527, HD 135344B, Elias
2-27;][]{grady13,christiaens14,stolker16,perez16}, and may
be excited by a secondary body or a massive planet
forming in-situ \citep{lin79,goldreich79,ward86}, by gas inflow
from an external envelope \citep{lesur15}, or by gravitational
instability (GI) \citep{toomre64, Gammie2001,
  durisen07}. In particular, $50 \%$ of Class I PP discs and $20 \%$
of Class II are believed to be gravitationally unstable
\citep{tobin13, mann15}, which suggests that spiral
waves certainly prevail during the early life of such objects. 

Recent observations of PP discs  in scattered light indicate that
micron-size dust is dispersed over a large
range of altitudes (typically 3 or 4 disc heightscale) in the vicinity
of spiral arms \citep{perrin09,benisty15}. In addition, 
direct mapping of gas emission and
velocity dispersion in CO reveal complicated structures within
the spiral wave \citep{christiaens14}.
Vertical circulation associated with
the waves
or turbulence arising from instabilities
\citep{bae16,riols17b}
might account
for the observations because these motions could
induce a vertical
diffusion of small dust particles. This mixing will also
impede dust
concentration and sedimentation, necessary ingredients in
planet formation \citep{chiang10}. 
Note that most disc models that describe the launching of spiral
waves by embedded
planets
\citep{zhu15,dong15} neglect the effect of dust circulation or mixing
when estimating the wave contrast in scattered light, 
possibly resulting in an overestimation of the planet mass. 

This paper is devoted to exploring the properties and origin of the
vertical
motions associated with 3D 
nonlinear spiral density waves. We focus especially on
the role of the disk thermodynamics and  
the context of GI turbulence.
{The 3D structure of spiral waves has been extensively studied in the framework
of linear theory \citep{loska86,lubow93,kory95,ogilvie98waves,
  lubow98}, whereas numerical simulations have only just
 started to explore their nonlinear behaviour.}  Hydraulic jumps
(i.e. the loss of hydrostatic balance
 behind a spiral shock) are one example of nonlinear effects known to
 produce a rapid vertical
 expansion of the gas and vortical flows (or `rolls') \citep{boley06}.
Vertical rolls may also be generated indirectly via waves'
deposition of heat and subsequent convective instability, as
demonstrated in some simulations of GI \citep{boley06b} 
and embedded planets \citep{lyra16}.
Finally, recent local
simulations of GI in \emph{convectively stable} discs by
\citet{shi10,riols17b} show that strong vertical motions
accompany spiral shocks on scales $\gtrsim H$. This
circulation exhibits a coherent but complex dynamics, with an
incompressible  and rotational part similar in strength to the
horizontal compressible motions. Other physics that may be relevant
but remains poorly understood includes
wave `breathing' (i.e. the interaction with the free surface), 
couplings with buoyancy motions (g-modes), and disk stratification
generally. Currently there is no general theory tying together these
diverse threads. This paper aims to be a step in such a
direction.

 We first analyse the vertical motions of spiral density waves, for
 different disc thermal stratifications. To that end, we combine
  linear axisymmetric calculations and PLUTO shearing box
 simulations of forced individual spiral waves.  
 In the case
 of a vertically sub-adiabatic (convectively stable) disk, a pair of 
large-scale coherent rolls emerge in the low-amplitude regime 
and we associate these features with g-modes excited alongside the density waves. 
 In the nonlinear regime the rolls are much more developed and,
importantly,  extend all the way to the midplane. The circulation
issues from the
 baroclinicity of the flow (misalignment between the pressure and
density gradients). When the
 wave-amplitude is sufficiently large, rolls emerge even for an
 adiabatic stratification with no vertical entropy gradient.  They
 arise from the shock wave structure itself, but remain marginal and
 rather shorted-lived.  

Second, we test the robustness of our result
 in less idealised gravitoturbulent disc simulations,
  where spiral density waves are excited naturally by
 GI. These simulations, characterising the early phase of PP discs,
 show that pairs of coherent poloidal rolls are created just above the
 spiral patterns and are supported by the background entropy gradient
 of the disc.  We discuss later the implications of these motions for
 astrophysical discs, in particular their relative 
 importance in vertical mixing, dust settlement, and dynamo action.

\section{Model and numerical setup}
\label{sec_model}
\subsection{{Governing equations}}
To study the spiral waves dynamics, we use a local Cartesian model of
the disc (Goldreich and Lynden-Bell 1965, Latter and Papaloizou
2017). 
In this model, the differential rotation is approximated locally by a
linear shear flow
 and a uniform rotation rate $\boldsymbol{\Omega}=\Omega \, \mathbf{e}_z$, with
shear rate $S=(3/2)\,\Omega$ for a Keplerian equilibrium. We denote $(x,y,z)$ as
the shearwise, streamwise and spanwise directions respectively, corresponding to the radial, azimuthal and vertical directions.  
We refer to the $(x,z)$ projections of vector fields as their poloidal
components. In most of the simulations, 
we assume that the gas orbiting around the central object is ideal,
its pressure $P$ 
and density $\rho$ related by $\gamma P=\rho c_s^2$, where $c_s$ is the
sound speed and $\gamma=5/3$ the ratio of specific heats (an
isothermal gas with $\gamma=1$ will be occasionally used, in
particular in section \ref{spiral_isoth}).  We also denote by
$S=c_V\ln({P}/\rho^{\gamma})$ the entropy
 of a fluid element with $c_V$ the heat capacity at constant
 volume. The evolution 
of density ${\rho}$, total velocity {field} $\mathbf{v}$, and pressure $P$ follows:
\begin{equation}
\dfrac{\partial \rho}{\partial t}+\nabla\cdot \left(\rho \mathbf{v}\right)=0,
\label{mass_eq}
\end{equation}
\begin{equation}
\frac{\partial{\mathbf{v}}}{\partial{t}}+\mathbf{v}\cdot\mathbf{\nabla
  v} +2\boldsymbol{\Omega}\times\mathbf{v} =-\nabla\Phi
  -\dfrac{\mathbf{\nabla}{P}}{\rho},
\label{ns_eq}
\end{equation}
\begin{equation}
\dfrac{\partial P}{\partial t}+\nabla\cdot (P\mathbf{v})
  =-P(\gamma-1)\nabla\cdot\mathbf{v}-\Lambda^{-}.
\label{int_energy_eq}
\end{equation}
In the shearing sheet model,  the total velocity field can be decomposed into a mean shear and a perturbation $\mathbf{u}$: 
\begin{equation}
\mathbf{v}=-\tfrac{3}{2} x\,  \mathbf{e}_y+\mathbf{u}.
\end{equation}
$\Phi$ is  the sum of the tidal gravitational potential induced by the
central object in the local frame $\Phi_c=\frac{1}{2}\Omega^2
z^2-\frac{3}{2}\Omega^2\,x^2$  plus an additional term, which can be
either an external potential $\Phi_\text{ext}$ (see Section
\ref{sec_spiralwaves}) or the gravitational potential induced by the
disc itself $\Phi_s$ (see Section \ref{sec_GI}) which satisfies the Poisson equation 
\begin{equation}
\mathbf{\nabla}^2\Phi_s = 4\pi G\rho.
\label{poisson_eq}
\end{equation}
 In the energy equation, we introduce  a cooling term $\Lambda^{-}$,
 necessary to simulate a gravito-turbulent disc in Section
 \ref{sec_GI}. We neglect  viscosity
 and thermal conductivity.  Finally, $\Omega^{-1}=1$ defines our unit
 of time and $H_0=1$ our 
unit of length. $H_0$ is the standard hydrostatic disc scale height 
defined as the ratio $c_{s_0}/\Omega$ where $c_{s_0}$ is the midplane sound speed. 

\subsection{Vorticity equation and baroclinicity}

To study the geometry of the flow associated with spiral waves, the
most useful quantity  is  the vorticity:  
\begin{equation}
\boldsymbol{\omega}=\nabla\times\mathbf{u}.
\end{equation}
In this paper, we mainly focus on the vortex dynamics in the
poloidal plane which is 
quantified by the $y$-component of $\boldsymbol{\omega}$. 
The equation governing the dynamics of $\omega_y$ in the rotating frame is 
\begin{equation}
\label{eq_vorticity}
\dfrac{D \omega_y}{Dt}= -\omega_y
(\nabla\cdot\mathbf{u})+\left(\left[\boldsymbol{\omega}+2\Omega\mathbf{e}_z\right]\cdot\nabla\right)u_y
+\frac{\left(\nabla \rho \times  \nabla P\right)_y}{\rho^2}
\end{equation}
where $D/Dt$ denotes the Lagrangian derivative.  On the right-hand
side, 
the different terms from left to right correspond to the stretching by
compressible motions, incompressible gradients,
 and baroclinicity. The last term is due to a misalignment of density
 and pressure gradient in the poloidal plane. If initially the gas is
 adiabatic with  $P\propto \rho^\gamma$ and does not dissipate energy
 into shocks 
during its evolution, then it remains barotropic. 
Note that the baroclinic term can be re-written as the cross product
of 
temperature and entropy gradients ${\nabla T \times  \nabla S}$. 
To produce baroclinic flows,  entropy has to be added to the system,
via shocks or other dissipative processes. 

\subsection{Background  equilibrium}
\label{bgd_equilibrium}
\begin{figure}
\centering
\includegraphics[width=\columnwidth]{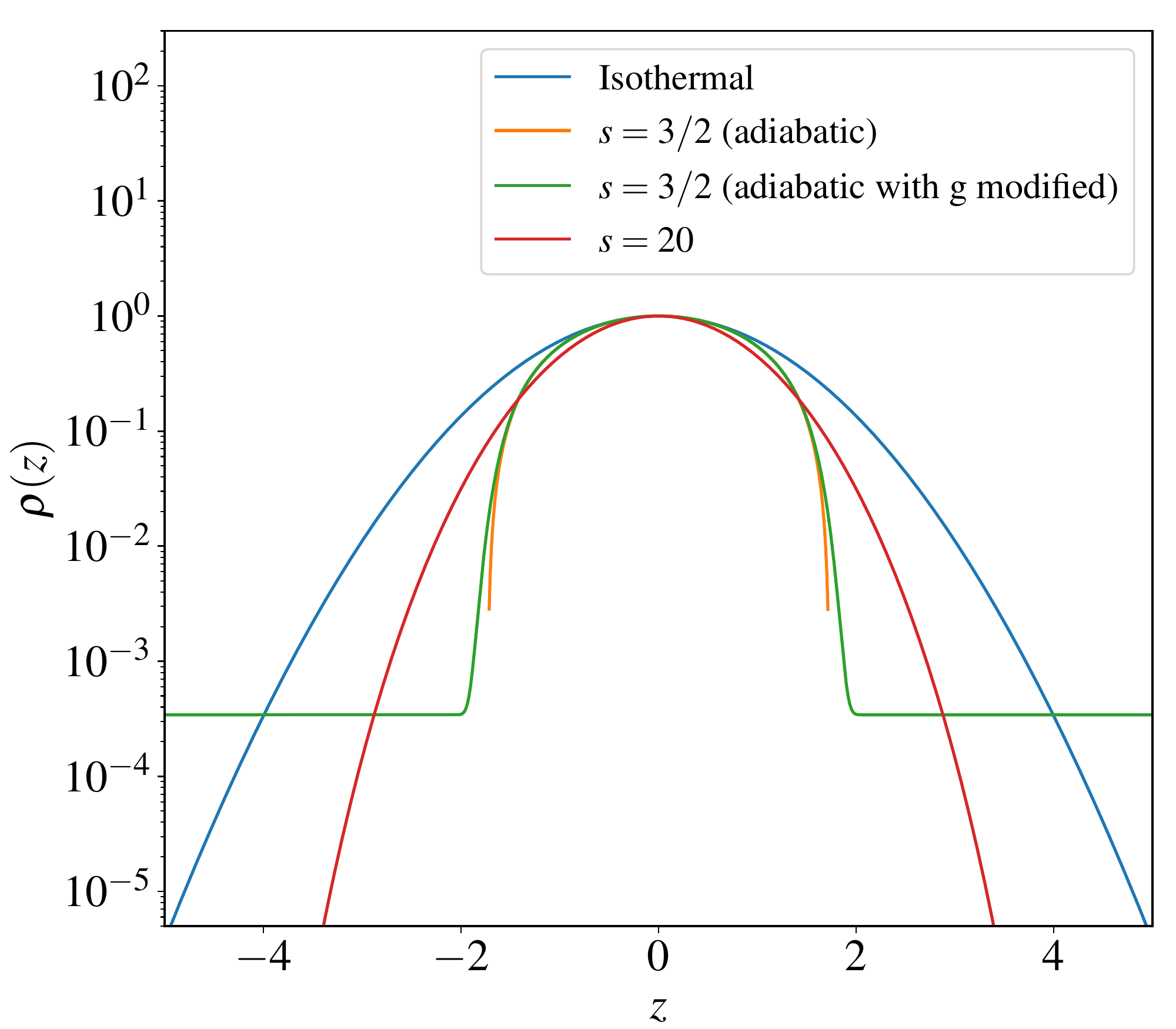}
 \caption{Density equilibrium profiles used in Section
   \ref{sec_aximode} and \ref{sec_spiralwaves}, serving as a
   background support for the  spiral density waves. The green curve
   corresponds to an adiabatic profile  with a modified gravity so
   that that the density converges toward a small constant value at
   large $z$. It is used in order to avoid a density floor in the adiabatic simulations of Section \ref{sec_spiralwaves}.} 
\label{fig_profiles}
\end{figure} 

In Section \ref{sec_aximode} and \ref{sec_spiralwaves}, {we neglect self-gravity and assume that}
waves propagate through a polytropic disc equilibrium, in which the
pressure is given by
\begin{equation}
P=K\rho^{1+1/s},   \quad \quad \text{with}  \quad  K=\dfrac{c_{s_0}^2}{\gamma \rho_0^{1/s}},
\end{equation}
where the subscript ``0" denotes the quantities in the
midplane. {We emphasize that this polytropic relation
  is adopted only as an initial equilibrium and
  does not account for the equation of state} The case
$\gamma=1$, $s=\infty$ corresponds to an isothermal atmosphere. An
adiabatic atmosphere corresponds to  $\gamma=1+1/s= 5/3$  (i.e
$s=3/2$), while for a general polytrope with stable stratification, we
have $s<3/2$. The entropy gradient is zero in the adiabatic case, while
in the polytropic case  it is related to the Brunt-V\"ais\"al\"a frequency $N^2$ through
\begin{equation} 
\dfrac{d s}{dz}=\dfrac{c_V \gamma N^2}{g(z)} = c_V(1+1/s-\gamma)\dfrac{d \log{\rho}}{dz}.
\end{equation}
Taking a general vertical gravity ${g}(z)$ and applying the vertical hydrostatic balance, the density profile of the equilibrium can be integrated analytically:
\begin{equation} 
\label{eq_density_profile}
\rho(z)=\rho_0\left(1-\dfrac{1/s}{(1+1/s)\Omega^2} \dfrac{\gamma}{H_0^2} \int^z_0 \mid {g}(z) \mid dz \right)^{s}.
\end{equation}
For almost all cases considered we use the thin disc approximation, for which the vertical gravity induced by the central object is
\begin{equation}
g(z)=-\Omega^2 z.
\end{equation}
Figure \ref{fig_profiles} shows some density profiles corresponding to
the isothermal case, to $s=3/2$, and to $s=20$.  Note that for an isothermal
equation of state, Eq.~(\ref{eq_density_profile}) does not apply and
the profile is the classical Gaussian $\propto \exp(-z^2/H_0^2)$.
For the other profiles, the density goes to 0 at a finite distance
$H_c=\sqrt{2(s+1)/\gamma} \, H_0$ from the midplane. 

\subsection{Numerical methods}
\label{numerical_methods}

In Section \ref{sec_aximode}, we linearise equations
(\ref{mass_eq})-(\ref{int_energy_eq}) and solve for the free axisymmetric
modes by using an eigensolver based on a Chebyshev collocation method
on a Gauss-Lobatto grid. This method results in a matrix eigenvalue
problem that can be solved using the QZ algorithm
\citep{golub96,boyd01}. Numerical convergence is guaranteed by
comparing eigenvalues at different grid resolutions and eliminating
the spurious ones.  We apply a  free surface boundary condition at the
vertical domain boundaries
$z=L_z/2$ and $-L_z/2$  with zero Lagrangian pressure,
 as in \citet[][hereafter KP95]{kory95}. 
We tested our solver by comparing eigenvalues and eigenfunctions with
those of KP95 and found very good agreement; in addition some of the
calculations 
were repeated with a shooting method similar to that of KP95 .   

In Section \ref{sec_spiralwaves} and \ref{sec_GI},  direct simulations
of equations (\ref{mass_eq})-(\ref{int_energy_eq}) are performed using
the  Godunov-based PLUTO code \citep{mignone2007}.  The numerical
scheme 
employs a conservative finite-volume method that solves the approximate Riemann problem at each inter-cell
boundary. Our simulations are  computed in a shearing box
 of size $(L_x,L_y,L_z)$  and resolution $(N_X,N_Y,N_Z)$. Note that because PLUTO conserves the total energy, the
heat equation Eq.(\ref{int_energy_eq}) is not solved directly. The code,
consequently, captures the irreversible heat produced by shocks due to
numerical diffusion, 
consistent with the  Rankine Hugoniot conditions.
Boundary conditions are periodic in $y$ and shear-periodic in $x$. In
the vertical direction  we use a standard outflow condition for the
velocity field but compute a hydrostatic balance in the ghost cells
for pressure.  As the problem studied varies from Section
\ref{sec_spiralwaves} to  Section \ref{sec_GI}, more details about box
size, resolution, 
initialization, and others parameters are given in the corresponding sections.

\section{Linear axisymmetric theory and roll structure}
\label{sec_aximode}

In order to set the scene, this section briefly 
revisits the linear theory of unforced
axisymmetric waves in a polytropic disc {without self-gravity} and highlight their vertical
structure. Though obviously not spiral
waves, they nevertheless encapsulate 
most of the physics we are interested in, especially when in the
tightly wound limit.  
Direct connections with simulated 
spiral density waves will be made in Section \ref{sec_spiralwaves}.

 \begin{figure}
\centering
\includegraphics[width=\columnwidth]{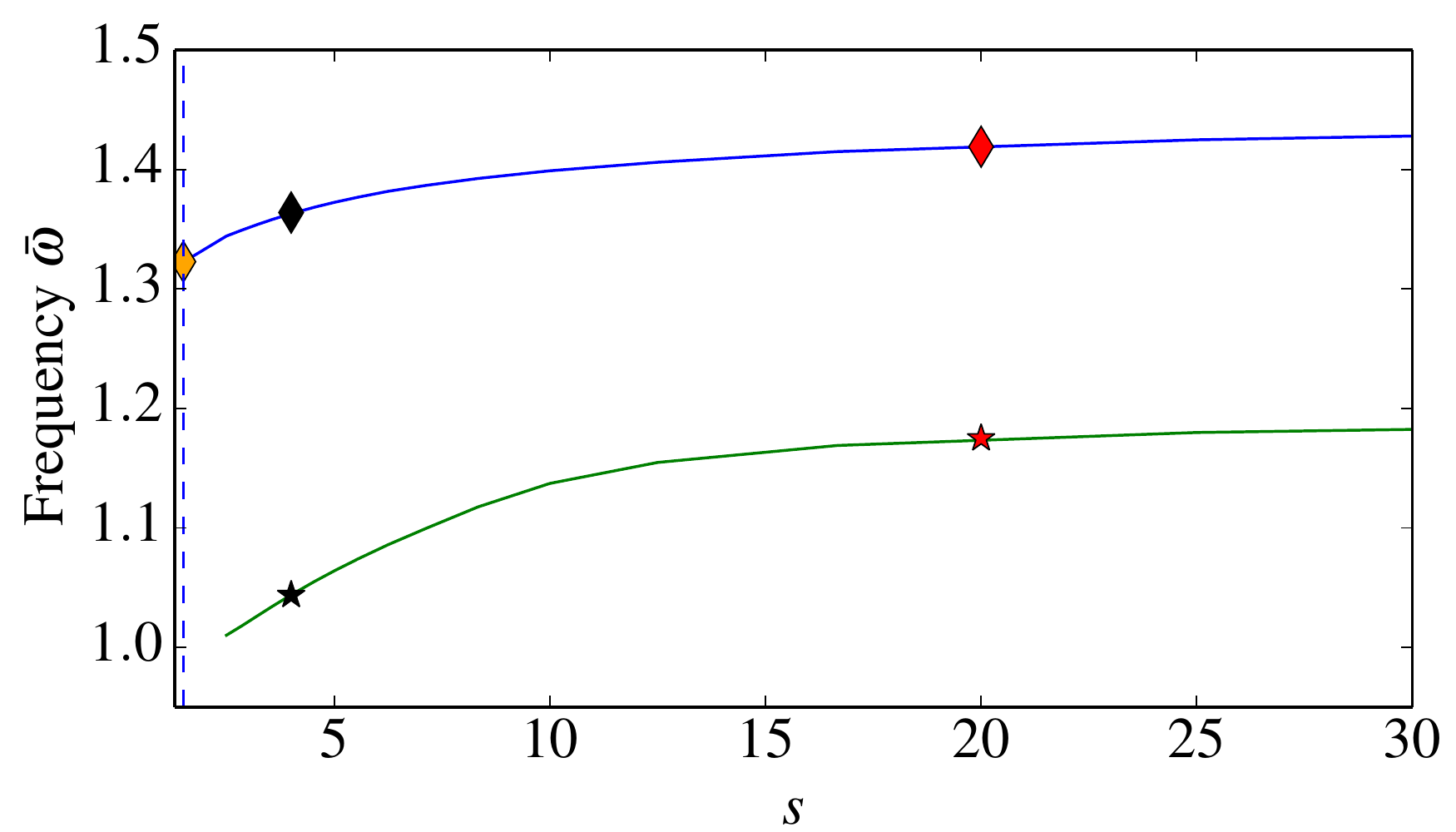}
\includegraphics[width=1.05\columnwidth]{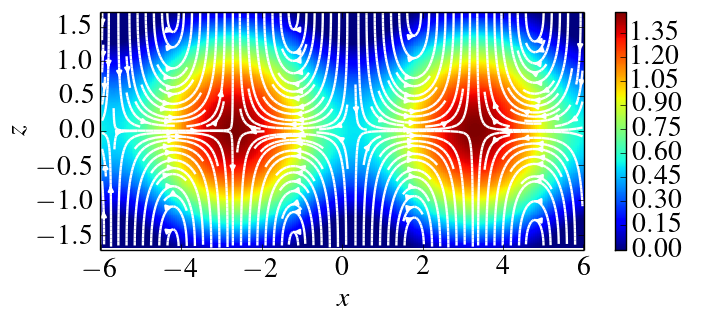}
 \caption{Top: Frequencies of the fundamental axisymmetric p-mode and
   g-mode $(n=0)$ as a function 
 of the polytropic index  $s$ and for $k_x= (\pi/3)
 H_0^{-1}$. Bottom:
 density and poloidal streamlines 
of the fundamental axisymmetric p-mode in an adiabatic atmosphere 
($s=3/2$) for the same $k_x$. The density shown here is $\rho_e+0.6\hat{\rho}$ }
\label{fig_linear1}
 \end{figure} 
   \begin{figure}
\centering
\includegraphics[width=1.03\columnwidth]{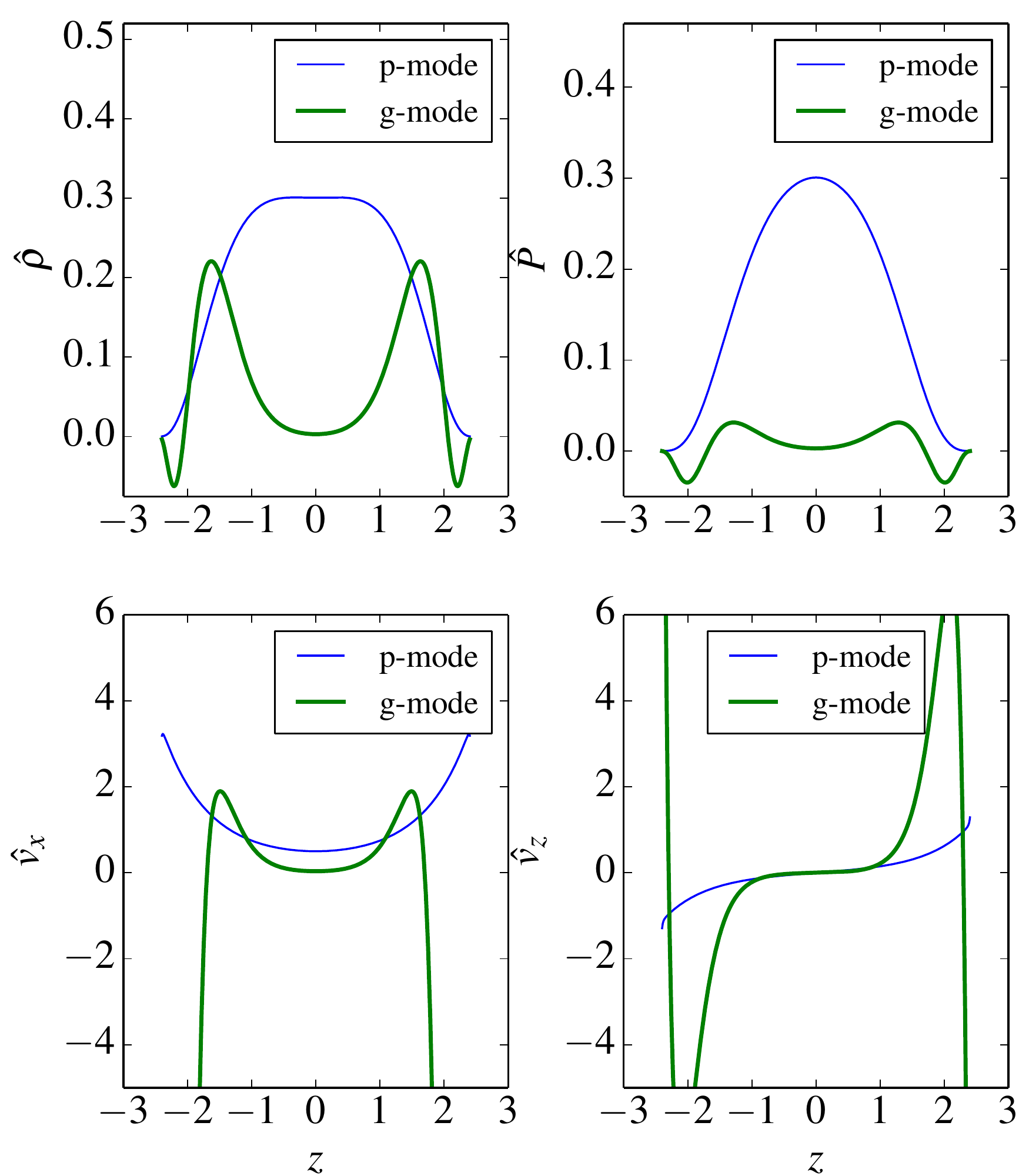}
 \caption{Eigenfunctions of the fundamental p-mode (blue) and g-mode
   (green) for $s=4$ and $k_x= (\pi/3) H_0^{-1}$. Top panels:
   perturbed density (left) and pressure (right). Bottom panels:
   perturbed radial (left) and vertical (right)
   velocities}
\label{fig_linear3}
 \end{figure} 
\begin{figure*}
\centering
\includegraphics[width=\textwidth]{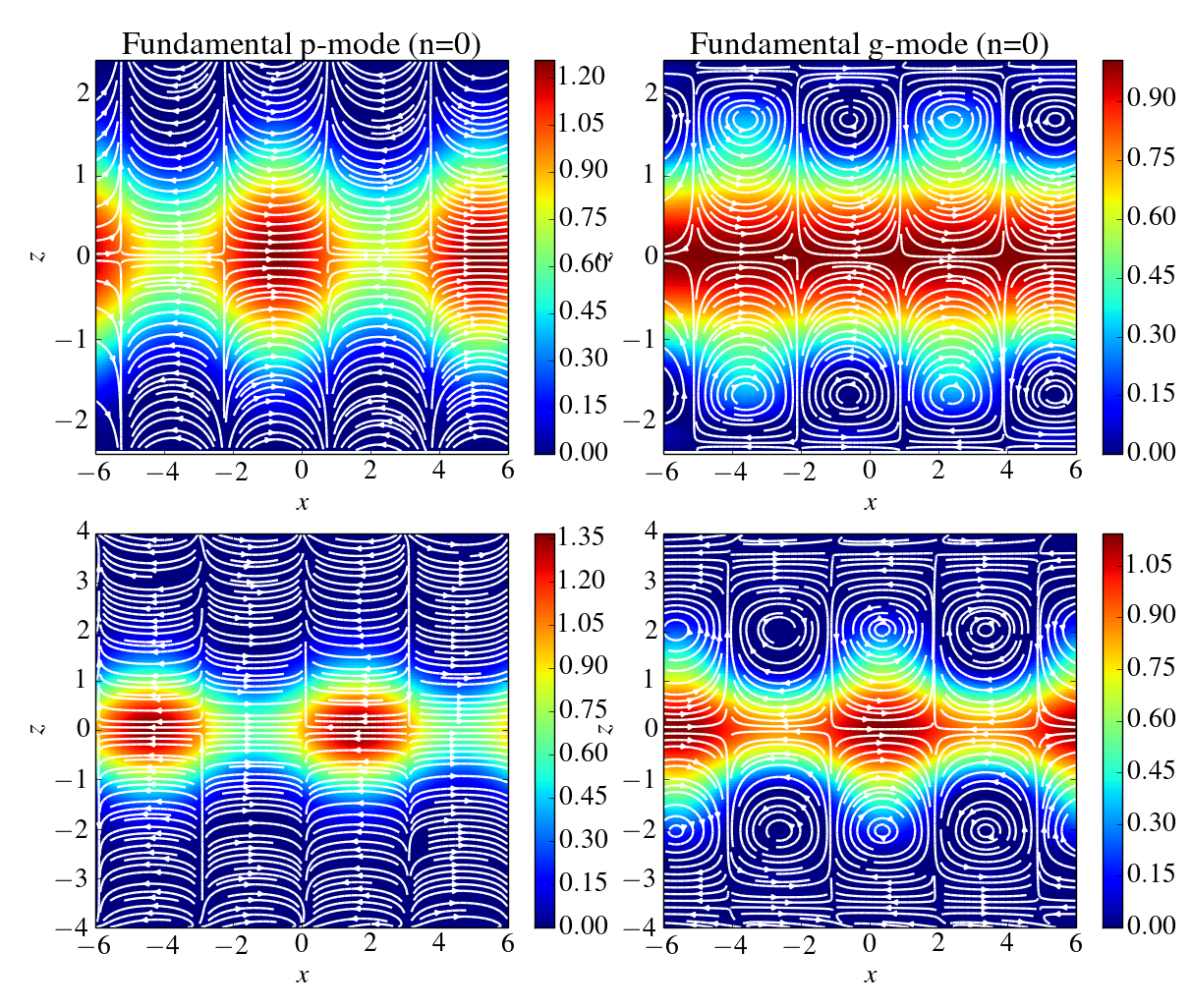}
 \caption{Density and poloidal streamlines of the fundamental axisymmetric p-mode (left) and g-mode (right) for $k_x= (\pi/3) H_0^{-1}$. Top panels are for $s=4$ while bottom panels are for $s=20$. The density shown here is $\rho_e+0.6\hat{\rho}$.}
\label{fig_linear4}
 \end{figure*} 

We denote by ${\rho}_e(z)$  and $P_e(z)$ the equilibrium profiles of 
density and pressure, defined in
 Section \ref{bgd_equilibrium}.  We add density, velocity, and pressure perturbations of the form 
$\left[\hat{\rho}(z),\,\hat{\mathbf{u}}(z),\hat{{P}}(z)\right]\,\exp(\text{i}k_xx-\text{i}\omega
t)$.  The densities (background and perturbations) are then normalised
by the midplane density $\rho_0$, velocities by the uniform sound speed
$c_{s_0}$, and pressure by $\rho_0 c_{s_0}^2$. Finally we introduce
the dimensionless wave frequency $\bar{\omega}=\omega/\Omega$ 
and wavelength $\bar{k}_x=k_xH_0$ with $H_0=c_{s_0}/ \Omega$.  
The linearised and normalized system of equations reads
\begin{align}
&-\text{i}\bar{\omega} \hat{\rho}+\text{i}\bar{k}_x\rho_e \hat{u}_x+ \dfrac{d}{d\bar{z}}(\rho_e \hat{u}_z)=0,\\
&\rho_e(-\text{i}\bar{\omega} \hat{u}_x-2 \hat{u}_y)=-\text{i}\bar{k}_x \hat{P},\\
&\rho_e(-\text{i}\bar{\omega} \hat{u}_y+\dfrac{1}{2}\hat{u}_x)=0,\\
&\rho_e(-\text{i}\bar{\omega} \hat{u}_z)=-\dfrac{d\hat{P}}{d\bar{z}}-\hat{\rho} \bar{z},\\
&-\text{i}\bar{\omega} \hat{P}+\text{i}\bar{k}_x P_e \hat{u}_x+ \dfrac{d}{d\bar{z}}(P_e \hat{u}_z)=-P_e (\gamma-1) (\dfrac{d \hat{u}_z }{d\bar{z}}+\text{i} \bar{k}_x \hat{u}_x).
\end{align}
Solutions to these equations have been obtained by KP95 and
\citet{ogilvie98waves}.
 The system supports three solution families:  one at ``low" frequency
with $\bar{\omega} <1$, which corresponds to inertial modes (r-modes), 
and two at ``high" frequency with $\bar{\omega} >1$, which correspond to
acoustic modes (p-modes) and buoyancy modes (g-modes). Each family is composed
of a countable set of branches, where each branch is labelled by an
integer $n$  characterising the mode vertical wavenumber.
The two fundamental even $n=0$ and odd $n=1$  p-modes are often
referred 
as f-modes or free surface oscillations, though they are also the 3D
manifestation of 2D density waves and 
their properties are very different from the stellar f-modes \citep[KP95, ][]{mamat10}\\

We focus on the $n=0$ fundamental p and g-modes for a wavenumber
$k_{x}=\pi/3$; thus the waves are radially large-scale. The 
 heat capacity ratio is set to $\gamma=5/3$.
The numerical domain extends from $z=0$ to
$z=L_z=H_c=\sqrt{2(s+1)/\gamma}$, the latter
corresponding to the disc surface beyond which $\rho_e$ and $P_e$
are precisely 0. The eigenfunctions are resolved by 300 points in the $z$
direction.  

Figure \ref{fig_linear1} (top) shows the angular
frequencies of the fundamental $n=0$  p-mode (blue curve, upper
branch) and g-mode (green curve, lower branch) as a function of the
polytropic index $s$.  We first analyse the p-mode in the adiabatic
case $s=3/2$ (yellow diamonds on the top
branch). Fig.~\ref{fig_linear1} (bottom) shows the total density
(background plus perturbations) and the poloidal streamlines
associated with the eigenmode. The streamlines show non-circulating
vertical motions, which can be interpreted as  a free surface
oscillation characteristic of the large-scale density wave. 
We should emphasise at this point that streamlines are not pathlines,
and so the fluid displacements associated with the waves
will be small on account of them being in the
(low-amplitude) linear regime.

 We next
plot the shape of the fundamental p-mode and g-mode for $s=4$
(black markers in  Fig.~\ref{fig_linear1}). Appearing in
Fig.~\ref{fig_linear3} are the normalized eigenfunctions for both modes
and in Fig.~\ref{fig_linear4} (top), the total density and poloidal
streamlines of the corresponding mode. Clearly, 
 the p-mode has the same shape as in the adiabatic case, while the 
g-mode exhibits roll structures located below the disc surface. 
Fig.~\ref{fig_linear3} (bottom) shows that the same behaviour is
obtained for larger $s=20$.  {We remind the reader that
  self gravity has been omitted in these calculations.
  We checked independently that these structures are
  also obtained when solving for
 the linear g-modes of a self-gravitating disc with Toomre $Q=1.6$}. 

In conclusion, low amplitude axisymmetric waves exhibit
vertical circulation (rolls) in the sub-adiabatic regime. 
These are most pronounced in the profiles of the (stable) g-modes,
whereas the modes associated with low-amplitude  density waves do not exhibit such vortical structures.

\section{Vertical structure of spiral waves forced by a potential}
In this section, we treat forced, non-axisymmetric,
and highly nonlinear waves, such as those excited by tidal
interactions or GI. For that purpose, we simulate in the shearing box
the behaviour of individual 3D spiral waves excited by an external
potential.

\label{sec_spiralwaves}
\subsection{Simulation setup and wave excitation}
\label{spiral_setup}
\begin{figure*}
\centering
\includegraphics[width=0.9\textwidth]{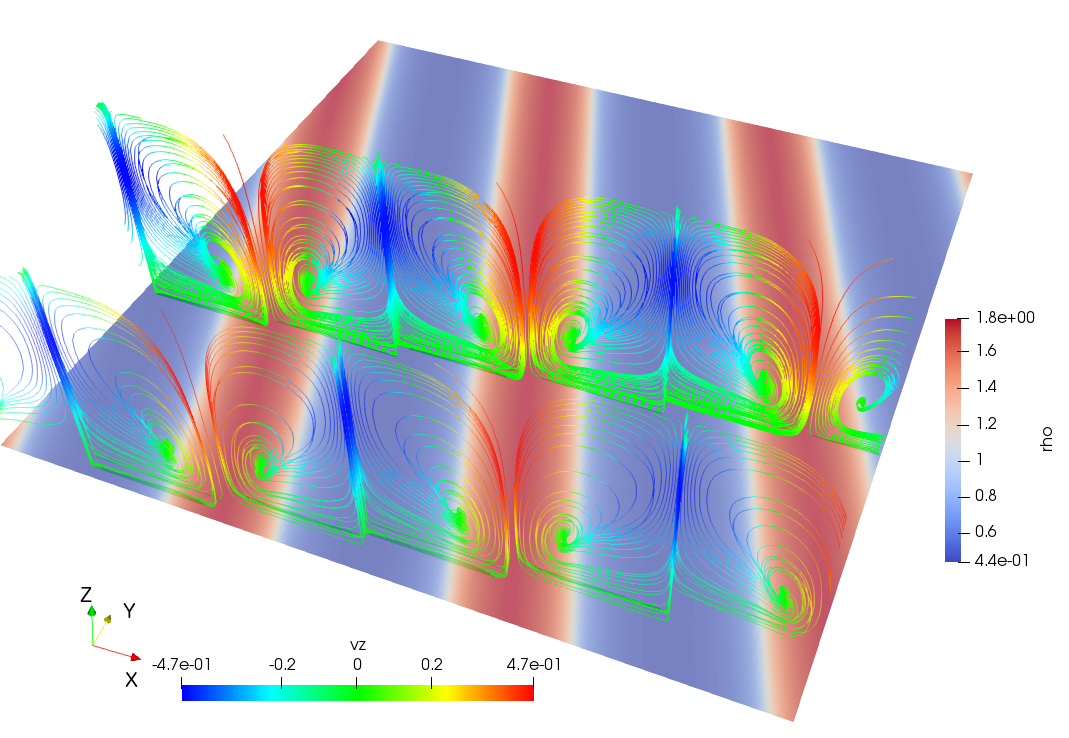}
 \caption{Top: 3D view of a spiral wave structure simulated with PLUTO
   in a polytropic disc atmosphere with $s=20$, and forcing amplitude
   $A=0.25$ (nonlinear regime). The wave density is represented in the
   midplane while the coloured lines are the streamlines  projected in
   the poloidal plane. The colour of the streamlines denote the
   vertical velocity.}
\label{fig_3Dview_s20}
 \end{figure*} 

 In order to characterize the fundamental motions associated with
 these waves, 
we restrict ourselves to a simple
 configuration, omitting self-gravity and retaining very
 basic thermodynamics (no
 cooling law). Our box spans $10H_0$ in the radial and azimuthal
 directions, and $\pm 3 H_0$ in the vertical direction. We start the
 simulations with different hydrostatic equilibria:
 isothermal, adiabatic, and polytropic (described in Section
 \ref{bgd_equilibrium}). Note that in the case of an adiabatic profile,
 $H_c\approx 1.73 H_0$ is small compared to the vertical extent of the
 simulation domain $z=\pm 3 H_0$, meaning that in most of the box the
 density is strictly zero, a numerical difficulty.
 One way to circumvent this
 problem is to impose a density floor beyond the altitudes $\pm H_c$.
 But doing so generates a strong shock associated with downfalling
 material. This shock produces a spurious entropy
 gradient capable of artificially producing vertical rolls. 
 A better alternative to the density floor is to use a slightly modified
 gravitational acceleration (in the adiabatic case only), such as
\begin{equation}
g(z)=-\Omega^{2} z \exp(-z^{16}/r_0^{16}),
\end{equation}
with $r_0=1.78 \, H_0$. The resulting density profile, shown in
Fig.~\ref{fig_profiles}
 is indiscernible  from the original (thin disk) profile for
 $z<1.73\, H_0$ and tends toward a fixed value of $ \sim 3\times
 10^{-4}$ at higher altitude. No spurious entropy is generated and
 yet the sharp density drop is well approximated at $z\simeq 1.73\,
 H_0$.

Once the equilibrium is given, the spiral waves are excited by an external potential of the form: 
\begin{equation}
\Phi_{\text{ext}}=A \,\cos\left[k_x(t)x+mk_{y_0} y\right],
\end{equation}
where $A$ is the amplitude of the forcing and
$k_x(t)=-k_{x_0}+Sk_{y_0}t$, with $k_{x_0}=2\pi/L_x$ and
$k_{y_0}=2\pi/L_y$ the fundamental wavenumbers of the numerical
domain.  We voluntarily impose no vertical dependence on the
potential, so that no vertical motions are directly forced. To a first
approximation, this potential reproduces the stirring of the gas by a 2D
self-gravitating potential as witnessed in simulations of
gravitoturbulence. 
Therefore, if vertical motions appear, they
result logically from the intrinsic spiral wave dynamics. 

The non-axisymmetric waves excited are not periodic
solutions of the fluid equations, but are rather dynamical. They are
first amplified by the potential during their leading phase  ($k_x(t)
k_y<0$, between $t=0$ and $t=S^{-1} L_y/L_x=2/3\, \Omega^{-1}$) and
then becomes trailing. After a few $\Omega^{-1}$, they reach a maximum
and then  decay because of the shear and numerical
dissipation.  An example
spiral wave is shown in 
Fig.~\ref{fig_3Dview_s20} at $t=2.5 \Omega^{-1}$ (during the late
trailing phase),  for a disc polytropic index $s=20$ and a forcing
amplitude $A=0.25$.  

\begin{figure*}
\centering
\includegraphics[width=\textwidth]{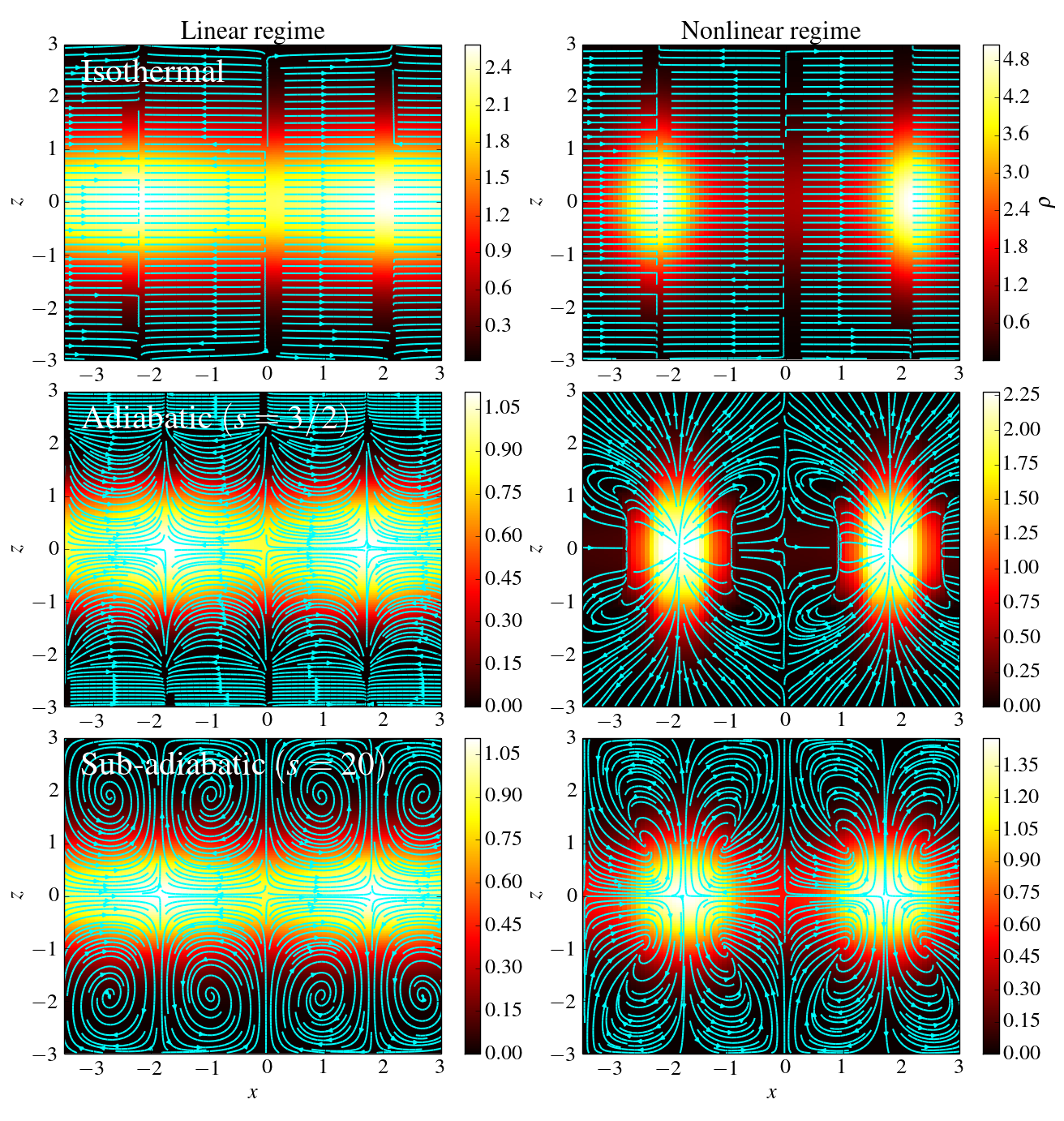}
 \caption{Poloidal structure of spiral density waves  in the
   isothermal  (top), adiabatic ($s=3/2$, middle) and sub-adiabatic
   atmosphere ($s=20$, bottom). The amplitude of the forcing in the
   left panels is $A=0.05$ (linear forcing). In the right panels, this
   amplitude is larger,  $A=0.4$, $A=1$, $A=0.25$ from top to bottom
   respectively (nonlinear forcing). 
  The colourmap indicates the density while the cyan curves are the
  streamlines in the poloidal plane.  
 Snapshots are taken at $t\simeq2.5\, \Omega^{-1}$. } 
\label{fig_streamlines}
\end{figure*} 
\begin{figure*}
\centering
\includegraphics[width=\textwidth]{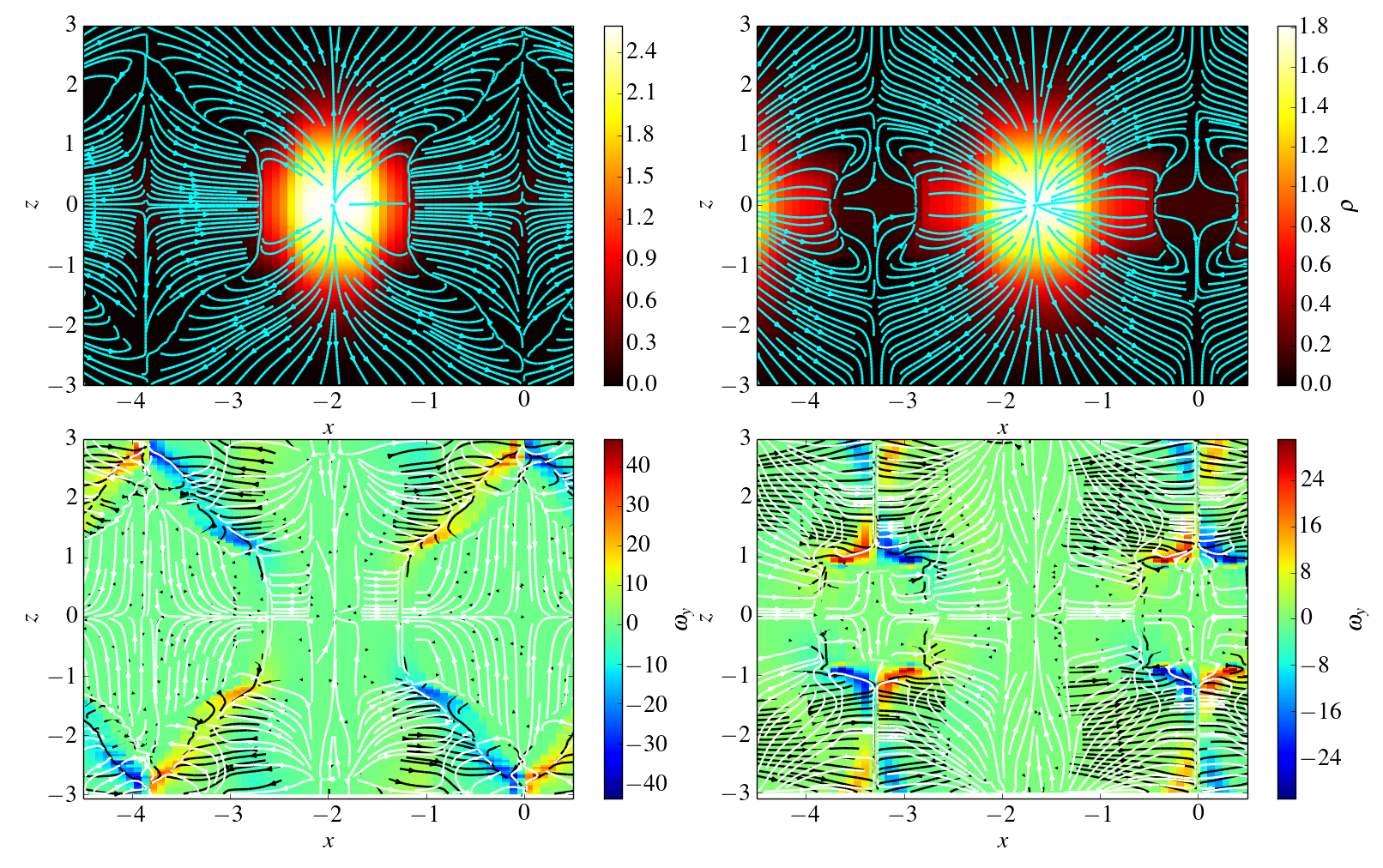}
 \caption{Top:  density and poloidal streamlines for $A=1$ in an
   adiabatic atmosphere ($s=3/2$). Bottom: vorticity in the $y$
   direction;  the arrows represent the direction of the temperature
   (white) and entropy (black) gradients.  Right and left  panels are
   at $t=2.4 \Omega^{-1}$ and  $t=2.7 \Omega^{-1}$ respectively.}
\label{fig_adiabatic_nonlinear}
 \end{figure*} 
 \begin{figure}
 \centering
 \includegraphics[width=1.07\columnwidth]{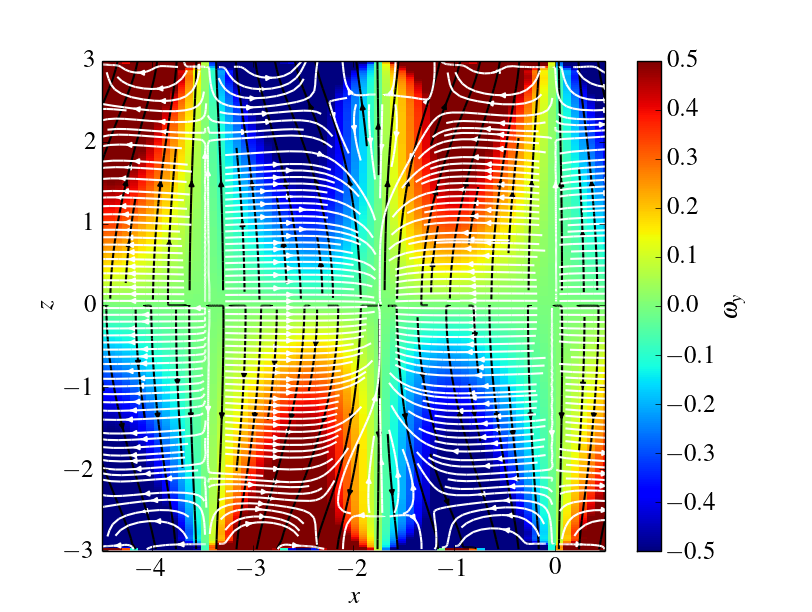}
\includegraphics[width=1.07\columnwidth]{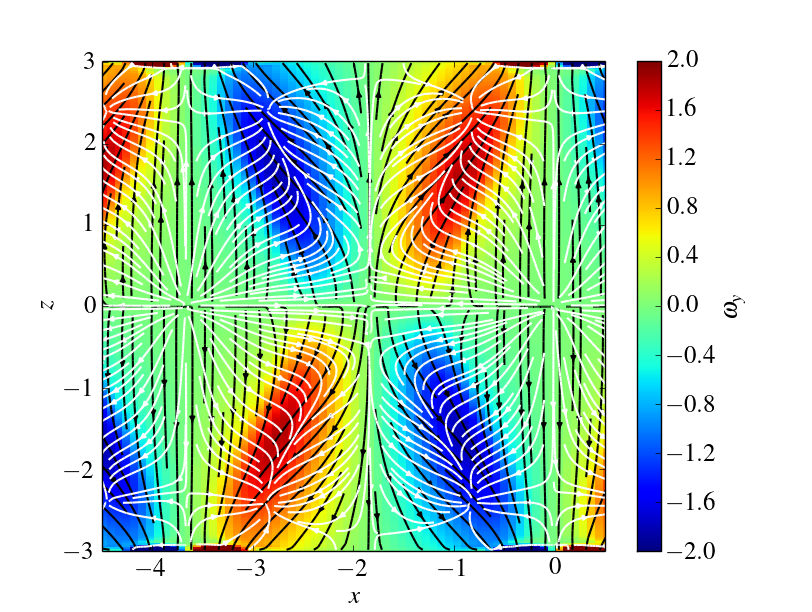}
 \caption{Vorticity in the $y$ direction for the case $s=20$ and
   $A=0.05$ (top) and $A=0.3$ (bottom); the snapshot is zoomed in on one spiral arm of Fig.~\ref{fig_streamlines}. The  arrows represent the direction of the temperature (white) and entropy (black) gradients. }
\label{fig_vorticity}
 \end{figure} 
 \begin{figure*}
\centering
\includegraphics[width=0.85\textwidth,trim=2.8cm 2.4cm 6cm 4.5cm, clip=true]{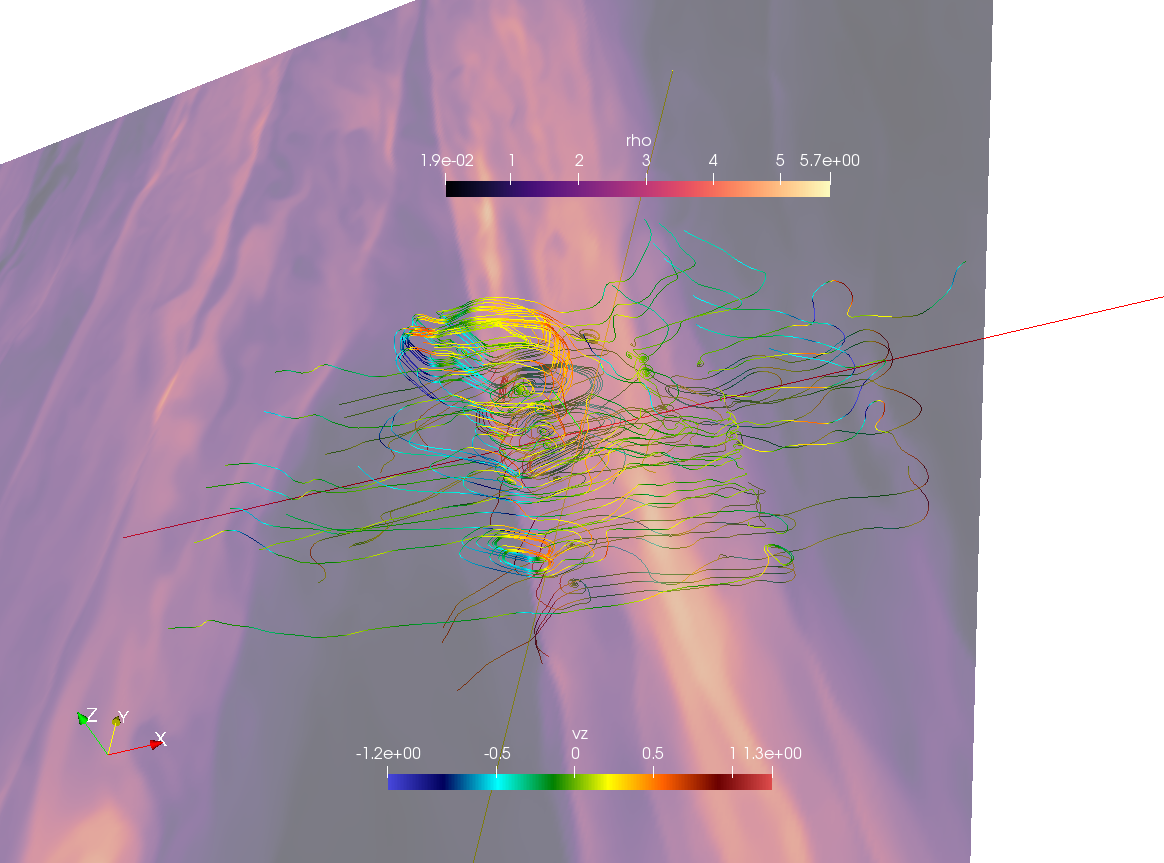}
\includegraphics[width=0.85\textwidth]{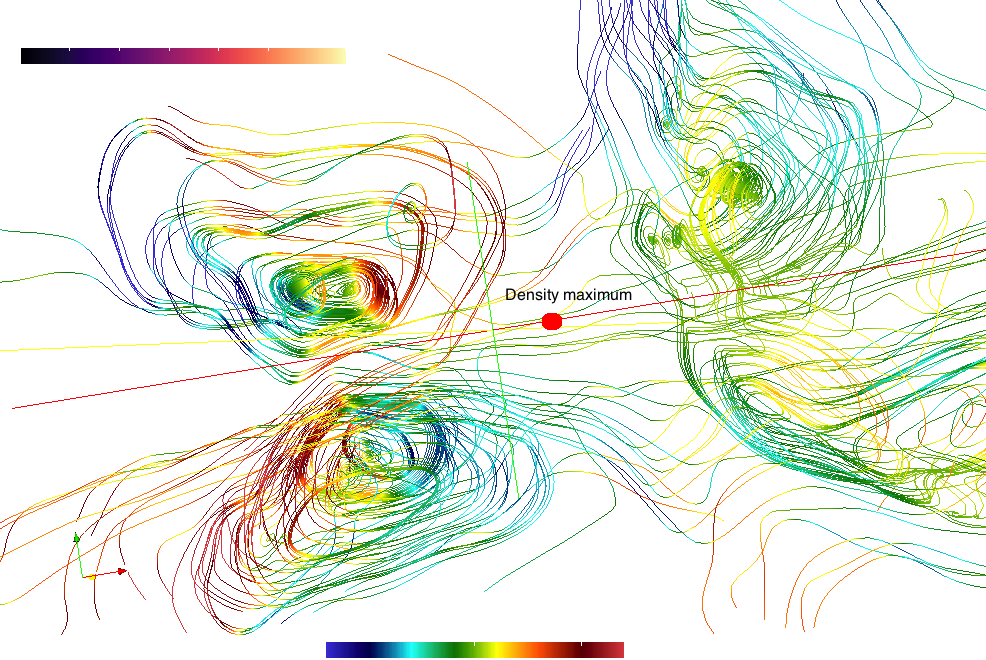}
 \caption{Streamlines associated with a spiral wave in a shearing box simulation of GI. The top panel is a view from the top (looking down to the midplane)  while the bottom panel is a poloidal cut (view from the disc midplane) with the angular momentum vector pointing upward.  The red line in the bottom panel indicates the midplane $z=0$. The colour of the streamlines represent in both cases the amplitude of the velocity component $v_z$}
\label{fig_GI}
 \end{figure*} 
 \begin{figure}
\centering
\includegraphics[width=\columnwidth]{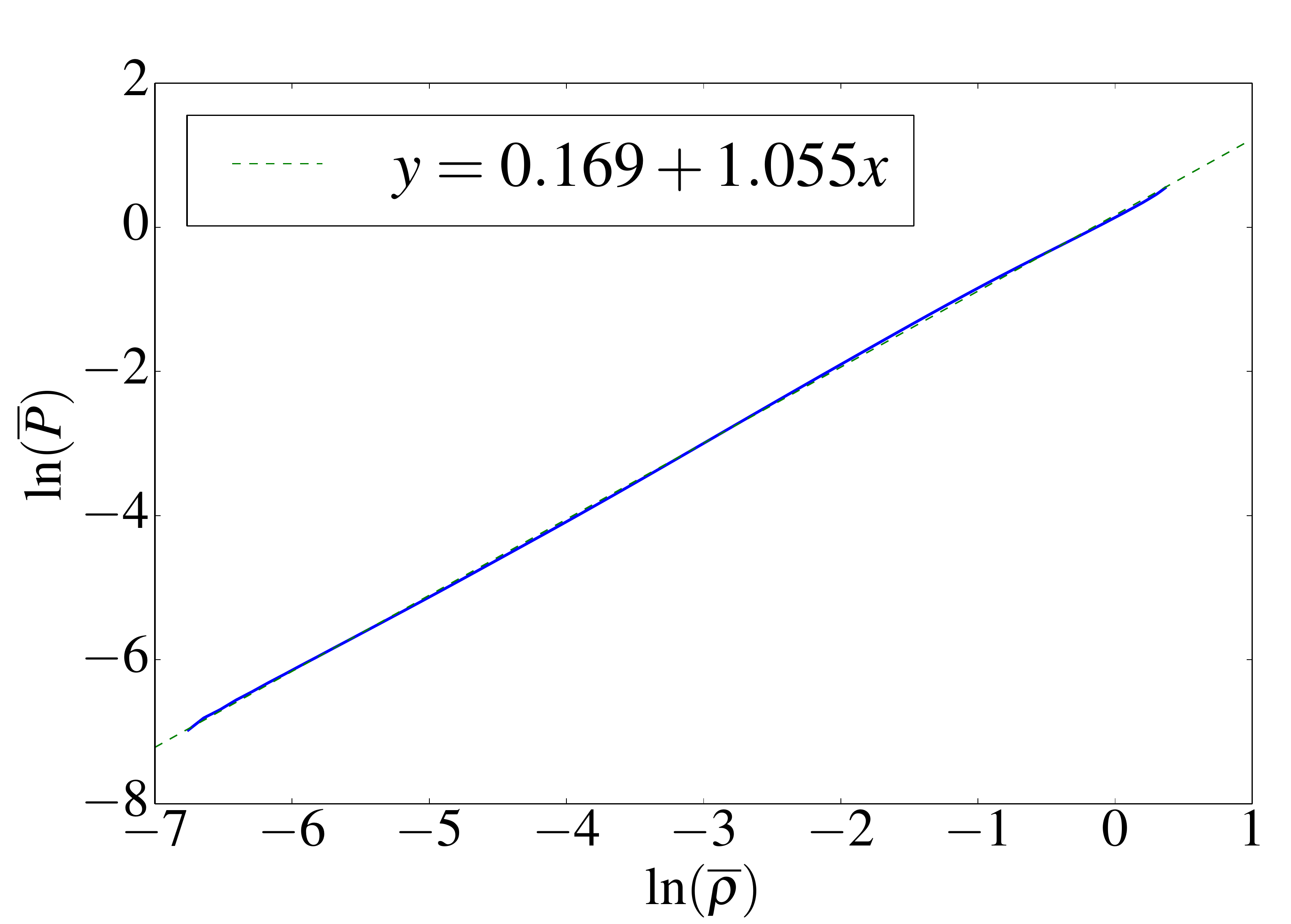}
 \caption{Disc stratification in the gravito-turbulent simulation of \citet{riols17b}. The plain blue curve represents the mean pressure profile in $z$  as a function of the mean density profile in $z$, calculated from the turbulent data (in log space). The dashed green curve is a linear fit of the plain curve. Clearly,  $\overline{P} \propto \overline{\rho} ^{1.05}$.}
\label{fig_GIstrat}
 \end{figure} 
\subsection{Linear regime}
\label{linear_spiral}

We first perform simulations of 3D spiral waves in the linear regime,
with a weak forcing $A=0.05$ and a resolution $N_X=N_Y=128$ and $N_Z=256$. 

\subsubsection{Isothermal}
\label{spiral_isoth}

When using an isothermal equation of state and an initial Gaussian
hydrostatic profile,  the forced spiral waves exhibit a relatively simple
vertical structure. Their poloidal motion consists of simple
compression/expansion and the streamlines remain purely horizontal, as
shown in the top left corner of Fig.~\ref{fig_streamlines}. (Higher
order p-mode harmonics (larger $n$) display non-negligible vertical displacements 
but these do not emerge naturally from a pure horizontal forcing, nor
could be classified as `density waves'.)
 If we leave
aside the non-axisymmetry, the mode forced here is similar to the
free two-dimensional acoustic–inertial mode $n=0$, which is the
first mode that becomes unstable to GI in isothermal disc \citep[see
Appendix B in ][]{riols17b}. 

\subsubsection{Adiabatic}

In contrast to the isothermal case, waves in an adiabatic 
stratification (with $s=3/2$) rapidly
develop vertical motions.  Between $t=0$
and $t=2\, \Omega^{-1}$,  the gas undergoes vertical ascending
motion at the radius where the density is maximum and a descending
motion where it is minimum (see centre-left panel of Figure
\ref{fig_streamlines}). The flow configuration resembles that
of the unforced axisymmetric wave (see  bottom panel of
Fig.~\ref{fig_linear1} for a comparison).  In the highly compressed
region $\nabla\cdot\mathbf{\delta u}<0$ and the pressure work is
positive. It induces a $\delta P>0$ with positive vertical gradient
$\nabla \delta P >0$, making the flow rise up.  On the other hand, in
the expanding region, $\nabla\cdot\mathbf{\delta u}>0$ and
the  pressure fluctuation is negative  but smaller in the corona. In
this case, the pressure difference  drives the flow toward the
midplane. It must be said, however, that it is
 inappropriate to designate the motion as
``circulating"  because the flow does not form closed loops and
turns parallel at large $z$.   

Ultimately, after $t=2\Omega^{-1}$ (not shown here), the topology of
the flow alters and the streamlines change their orientation because
the gas expands in the densest regions. In the bulk of the disc
($z<H_0$), the vertical motions become oriented toward the
midplane. All streamlines converge to the same point which corresponds to
the density minimum. We checked that no vortical structure appears
during the late trailing  phase of the linear wave.  

\subsubsection{Sub-Adiabatic}
\label{sub_adiabatic}

Fig.~\ref{fig_streamlines} (bottom left) shows the poloidal
streamlines for $s=20$ which corresponds to a sub-adiabatic
stratification. This profile mimics a  disc in which the corona is
heated by an external source or by turbulent motions rather than
pressure work.  As we will show later in Section \ref{sec_GI}, it
adequately describes simulations of GI turbulent discs, which usually show flat
temperature profiles. 

In this regime, we found that in the linear
regime, spiral waves exhibit large-scale poloidal rolls,  orbiting or
spiralling toward a central point.   Each density maximum is
surrounded by four counter-rotating  cells, symmetric about the
midplane. These coherent motions are maintained for $\sim 1\,
\Omega^{-1}$.
By scanning other polytropic indices, we found that proper rolls  (with
closed streamlines) also emerge for smaller $s$.   

The vortical structures are similar to those exhibited by the unforced linear
axisymmetric g-modes ($m=0$, see Fig.~\ref{fig_linear4} for
comparison).   Note however that the density maxima in the simulated
forced waves are located at the intersections of the rolls, but not in
the unforced (free) linear axisymmetric calculation. 
Our explanation is that the
gravitational forcing excites both the fundamental p-mode (density waves) and
the fundamental g-mode. The density patterns are mainly due to the p-mode
while the velocity components above the midplane are the expression of
the  g-mode. These two superimposed modes possibly
interact weakly each other via nonlinear terms,  providing the observed
shape. The phase lag may result from this weak
coupling. The physical origin of the
rolls  is discussed in Section \ref{rolls_origin}.  

\subsection{Nonlinear regime}
\label{nonlinear_spiral}

We now explore strongly nonlinear waves, for which the
amplitude $A$ is adjusted in order to obtain a local Mach number $u/c_s$
larger than 1.  We keep the same resolution as in the previous
section.   

In the pure isothermal case, increasing the amplitude of the forcing
potential does not help the formation of vertical structures.  For an amplitude
$A\simeq 0.4$, Fig.~\ref{fig_streamlines} (top right) shows that the
streamlines, projected in the poloidal plane, remains horizontal.  

In the adiabatic atmosphere we examine two forcing amplitudes:
 $A=0.4$  and $A=1$.
Figure \ref{fig_streamlines} (centre-right panel)
reveals that for a large amplitude $A=1$,  the vertical wave structure
at $t=2.5\Omega^{-1}$ is more complex than in the linear regime.  A
pair of rolls is produced  and take the form of butterfly wings.
However, the latter remain relatively short-lived and are less
developed than in the linear sub-adiabatic
regime. Fig.~\ref{fig_adiabatic_nonlinear} (top panels) shows the
streamlines before and after $t=2.5\Omega^{-1}$. As the flow
rises up during the early trailing phase (left panel), a shock forms
and deflects the streamline, producing
vorticity (see next section). Later (right panel), the shock has been
advected by the vertical circulation and almost disappears. At a
smaller forcing amplitude  $A=0.4$, corresponding approximatively to the
transonic regime, the same behaviour is observed, but only marginally and the
vortical motions generated are much less pronounced. 

Finally, in the case of a subadiabatic polytrope with $s=20$ and 
a forcing of $A=0.3$, the spiral wave is characterised by
well-developed vertical cells as shown in Figure
\ref{fig_streamlines} (bottom right).
They are much stronger than in the nonlinear
adiabatic case and expand further away into the corona. A weak shock is
present and tends to  bend the streamlines,  preventing the rolls from
being circular. It is important to note that, unlike the linear
regime, the rolls extend all the way to the midplane and are probably
much more efficient at mixing disc material throughout the vertical
column. The 3D
view in Fig.~\ref{fig_3Dview_s20} shows that the poloidal wave pattern
in the nonlinear regime maintains a translational symmetry along the
density wake.  

\subsection{Physical origin of the rolls}
\label{rolls_origin}

In the adiabatic case, rolls appear only if the local Mach number of
the wave exceeds unity.  Figure  \ref{fig_adiabatic_nonlinear}  shows
the poloidal streamlines (top panels) at two successive times, $t=2.4$
and $t=2.7 \, \Omega^{-1}$, when the forcing is $A=1$.  In the left panel, a
shock is clearly visible and separates the expanding region (where the
flow rises up) and the compressed region (where the flow converges
toward the density maximum). Observe that the shock is not fixed but
advected by the ascending flow. Therefore, the postshock region (or
downstream velocity)  is located above, where the gas expands.   The
bottom panels of Fig.~\ref{fig_adiabatic_nonlinear}  shows the
vorticity and orientation of the entropy (black lines) and temperature
(white lines) gradients. Vorticity is produced across the
entropy jump at the shock interface.   The misalignment between
entropy and temperature gradients in the post-shock region contributes
also to the vorticity production in this region via the baroclinic
term $\mathbf{\nabla} S \times \mathbf{\nabla} T$ (see
Eq.~\ref{eq_vorticity}). The other terms  in Eq.~\ref{eq_vorticity}
are subdominant during this phase. In summary, rolls motions are
driven by a nonlinear baroclinic effect,  mediated by the entropy
production across the shock wave.    
 
In the sub-adiabatic case, the mechanism 
differs, since rolls emerge even in the linear theory; we argued
in Section \ref{sub_adiabatic} that they are the signatures of 
large-scale g-modes excited alongside the density wave. But what is
the physical mechanism driving these rolls?  Fig.~\ref{fig_vorticity} (top) shows
that for a small forcing amplitude $A=0.05$,  vorticity is concentrated along
butterfly ``wings" where entropy and temperature gradient are
orthogonal. The poloidal rolls are then clearly produced through the
baroclinic term. Unlike the adiabatic case, a vertical entropy
gradient is already present, inherited from the background
stratification.  No shocks are required.  The temperature
gradient has a strong radial component, resulting from the pressure
work due to wave compression. Fig.~\ref{fig_vorticity} (bottom) shows
that the same patterns are seen in the nonlinear regime,  although the
temperature gradients are less horizontal, in particular near the
shock region.  Additional vorticity is provided by the entropy jump
across the discontinuity and tends to deflect the streamlines, in a
manner similar to the adiabatic case.

\section{Spiral waves in gravito-turbulent discs}
\label{sec_GI}

In this section, we study the vertical dynamics of spiral density
waves excited by GI. Our goal is to check if the
vertical motions described in the previous section are robust and can be obtained
in a turbulent background with cooling typical of young PP discs. 

\subsection{Initial condition and simulation setup}

{To begin, we performed a  turbulent disc simulation
  with GI.  The simulation is initialized from a 3D gravito-turbulent
  state obtained from \citet{riols17b}}  (the run is labelled PL20-512
in the corresponding paper).  The numerical methods and simulation setup
are detailed in the related paper. We remind the reader that self-gravity
is computed by solving the Poisson equation numerically
(Eq.~\ref{poisson_eq}). 
To solve this equation, we perform a decomposition of the density
and potential in 2D Fourier space ($k_x$,$k_y$) for each slice in
$z$. The boundary conditions in the vertical direction are
non-periodic and given in \citet{riols17b}.  {The vertical domain
extends from  $-3H_0$ to $3H_0$}  and the horizontal box size
is $L_x=L_y=20\,H_0$. The resolution is $512\times512\times64$, so
that one scale-height is resolved by $\sim 26$ points. 
To make a steady gravito-turbulent state possible,  we introduced
a cooling  law $\Lambda^{-}=P/\tau_c$  in the internal energy equation
(\ref{int_energy_eq}) where $\tau_c=20\,\Omega^{-1}$ is a typical
timescale referred to as the `cooling time'. This prescription is not
necessarily realistic but allows us to control the rate of energy loss
via a single parameter.  

\subsection{Spiral waves and vertical rolls}

The gravito-turbulent simulation describes 
a flow composed of large-scale spiral density waves, small-scale incompressible waves, and
clumpy structures. We focus here on the dynamics around one spiral
wave chosen randomly within the turbulent flow. Fig \ref{fig_GI}
shows the 3D topology of the streamlines around this spiral
wave. The density wake can be seen in the coloured rendering of
the density $\rho$ in the top panel, which appears bright and pale when high, and dark
in more evacuated regions. Superimposed upon the wake are the flow
streamlines coloured according to the magnitude (and sign) of $v_z$.
The bottom panel is a poloidal cut of the same wake, with the red straight line
indicating
the midplane and the green straight line the vertical axis. The green
axis cuts through the centre of the density wake in the upper panel.
 
In the bottom panel, we can clearly distinguish on the left of the density wake two
(somewhat tangled) large rolls, with streamlines forming closed
loops. The sign of $v_z$ indicates the the upper roll rotates
counter-clockwise, while the lower roll rotates clockwise.
Both are of scale
$H$ and symmetric (on average) about the midplane.  On the right of the density wake,
the flow is much more disorganised but again one can discern
two roll-like structures; they
are fainter and rotating in the opposite direction. 
Overall, the streamlines form a pattern similar to
those depicted in  Fig.~\ref{fig_3Dview_s20} and
Fig.~\ref{fig_streamlines} (bottom). They are however more disorganised,
somewhat understandably, as they emerge from a turbulent
background.

To make a comparison with our simple atmosphere model of Section
\ref{sec_spiralwaves}, we estimate  the typical index $s$ measuring
the disc stratification in our GI simulations.  Figure
\ref{fig_GIstrat} shows the mean vertical pressure profile
$\overline{P}$ as a function of the mean density profile
$\overline{\rho}$,  in logarithm scale. Each profile  is averaged over
time, $x$ and $y$. The result shows a clear linear trend, suggesting
that the polytropic model is an excellent fit to the mean
gravito-turbulent vertical structure.  The slope of the curve gives
$1+1/s \simeq  1.05$, i.e $s\simeq20$,  which is the polytropic  index
used in Section \ref{sec_spiralwaves}. Therefore, there is clear
indication that the vertical rolls are driven by the vertical entropy
gradient naturally generated by the turbulent motions in these
discs. However, we emphasize that this gradient might be sensitive
to the cooling law and numerical details of the simulations. It is
probably an under-estimate, given that a real disc is usually irradiated
by FUV radiation, X-rays, and cosmic rays, physics that is not taken into account
in our simulations.  

Finally, we checked that these structures appear in other spiral arms
and occur frequently in the simulation. {We point out
  that the roll structures do not survive more than a few
  $\Omega^{-1}$, 
and die once their parent density waves break down}. The rolls motions are transonic
and represent a large fraction of the r.m.s velocity. They account for
most of the vertical kinetic energy in the disc corona and are clearly
responsible for the increase of both the r.m.s radial and vertical
velocity with altitude (see Fig . 2 of \citet{riols17b}). Note that in
addition to the  large scale rolls, a small-scale vertical component
arising from a parametric instability has been also  identified in
this simulation  \citep[see][]{riols17b}.

\section{Discussion and conclusions}

In summary, we have shown using shearing-box simulations that 
large-scale spiral waves in astrophysical discs exhibit vertical and
vortical gas motions of size comparable to the disc height-scale. For
an adiabatic atmosphere $1+1/s=\gamma=5/3$,  these structures appear
if the wave achieves shock amplitudes, but they are rather difficult to
produce and have a short lifetime.  When the polytropic index is lower
($1+1/s<\gamma$), vortical structures are much easier to trigger:
they appear  even in the subsonic regime and live longer. 
In this case, the flow is characterised by four counter-rotating
cells, symmetric about the midplane, potentially transporting material from 
overdense regions towards the surface. 
These structures are enhanced in the nonlinear regime
and penetrate to the midplane.  Finally, we found that roll
motions are ubiquitous in gravito-turbulent disc simulations and
are clearly related to the spiral wave dynamics induced by the
gravitational instability. They appear to resist the
small-scale turbulent background characterising these types of flows
(see \citet{riols17b}).  

We investigated the origin of these motions and found that they issue
from the baroclinicity possible in the
system. The effect is enhanced if the disc is sub-adiabatic
(stably stratified) with a large scale (stable) entropy gradient. In contrast,
rolls in adiabatic disks requires
sufficiently large wave amplitudes and associated shocks.
 Our semi-analytic and linear
axisymmetric calculations  reveal that the rolls are associated with
the fundamental (buoyancy) g-mode, which appears to emerge alongside the
density wave. The alternative mechanisms mentioned in the introduction
(vertical breathing/splashing, convection) might be also relevant for
the spiral wave dynamics but are not essential in the production
of rolls of typical size $\gtrsim H$.  

Our results have several implications for young protoplanetary disks especially.
Poloidal rolls may convey small grains (with stopping times much less than
$\Omega^{-1}$) as far as the corona
 if the associated
density waves are of sufficient amplitude and live sufficiently long.
Obviously, this physics could bear directly on dust concentration
and sedimentation processes in those PP discs subject to GI or
disturbances originating from embedded planets and binary companions.
In particular, vertical circulation may interfere with dust
agglomeration in gravitoturbulence. Current simulations
of GI-induced filamentary structure are 2D \citep{gibbons12,gibbons14,
  gibbons15, shi16} and cannot describe the 3D vertical
circulation revealed here.  Quite separately, dust can also be
mixed in the upper atmosphere by a parametric instability attacking
the spiral
waves themselves
\citep{bae16,riols17b}, and this could also affect the
agglomeration process.  

The roll motions could have an indirect impact on the scattered
infra-red luminosity measured from observations.
If the vertical lofting is efficient, dust may
settle above the spiral patterns at the disc surface, 
altering its emission properties. In discs with an
embedded planet this could bear on the estimation
of the density contrast of the spiral arm and the subsequent inference
of the planet's mass \citep[see][for more details on the original
problem]{juhasz15}.  
{We, however, emphasize that there are two main caveats to this effect}. {First there is generally a relative azimuthal velocity between dust particles and the spiral wave 
front. In that case,} the lift will be significant if the time dust spends within the spiral structure is comparable to the rolls' turnover time. This is probably true in the co-rotation region of the spiral
wave (near the planet) but less obvious further away where the
relative azimuthal velocity between dust particles and the wave
front is significant. In gravitoturbulence, this issue is probably less important
since spiral waves are almost co-rotating with the disc and are
excited almost uniformly in the unstable region.
{Second, we stress that streamlines are different than
  pathlines, since the spiral waves propagate radially. For instance,
  in the linear regime, the motion of a test particle does not follow
  the vertical rolls and by definition, has only a small displacement. Thus,
  the wave needs to be sufficiently strong for the dust to rise.} To quantify these
different effects and make more accurate predictions, simulations of
the dust-gas interaction in the vicinity of spiral waves or in 3D gravitoturbulence
need to be performed and will be the object of a future
paper.

A final application of this result is to the question of
large-scale magnetic field generation in astrophysical discs
generally. 
Recent simulations by
\citet{riols17c} revealed that the spiral wave dynamics induced by GI
can act as a powerful dynamo. The four counter-rotating roll motions
identified in this study  (combined with the shear) provides
large-scale helicity and hence may efficiently amplify a seed magnetic
field. In fact, the velocity structures are similar to the Ponomarenko or
Roberts flows, which are known to provide kinematic dynamo action. One
could then imagine a large-scale dynamo based on the successive action
of spiral waves, although further work is required to understand if
such a dynamo could work in gravito-turbulent discs, especially in the
presence of the non-ideal MHD regimes relevant for PP disks.
  
\section*{Acknowledgements}

The authors would like to thank the anonymous referee for a set of
helpful comments and suggestions, and Geoffroy Lesur and Gordon
Ogilvie for advice. This work is partly 
funded by STFC grant ST/L000636/1.

\bibliographystyle{mnras}
\bibliography{refs} 

\appendix

\label{lastpage}
\end{document}